\documentclass[aps,prl,twocolumn,superscriptaddress,showpacs]{revtex4}

\usepackage{graphicx}
\usepackage{amsmath}
\usepackage{color}

\begin{document}

\title{Sharp dark-mode resonances in planar metamaterials with broken structural symmetry}

\author{V. A. Fedotov}
\email{n.i.zheludev@soton.ac.uk}
\affiliation{Optoelectronics Research Centre, University of Southampton, SO17 1BJ, UK}

\author{M. Rose}
\affiliation{Optoelectronics Research Centre, University of Southampton, SO17 1BJ, UK}

\author{S. L. Prosvirnin}
\affiliation{Institute of Radio Astronomy, National Academy of
Sciences of Ukraine, Kharkov, 61002, Ukraine}

\author{N. Papasimakis}
\affiliation{Optoelectronics Research Centre, University of Southampton, SO17 1BJ, UK}

\author{N. I. Zheludev}
\homepage{www.nanophotonics.org.uk}
\affiliation{Optoelectronics Research Centre, University of Southampton, SO17 1BJ, UK}

\date{\today}

\begin{abstract}
We report that resonant response with a very high
quality factor can be achieved in a planar metamaterial by
introducing symmetry breaking in the shape of its structural
elements, which enables excitation of dark modes, i.e. modes that are weakly coupled to free space.
\end{abstract}


\maketitle

Metamaterials research has attracted tremendous amount of
attention in the recent years. The interest is mainly driven by
the opportunity of achieving new electromagnetic properties, some
with no analog in naturally available materials. Extraordinary
transmission \cite{Ebessen}, artificial magnetism and negative
refraction \cite{NR}, invisible metal \cite{FS}, magnetic mirror
\cite{Mirror2}, asymmetric transmission \cite{Asymmetry}
and cloaking \cite{Cloak} are just few examples of the new
phenomena emerged from the development of artificially structured
matter.

The exotic and often dramatic physics predicted for metamaterials
is underpinned by the resonant nature of their response and
therefore achieving resonances with high quality factors is
essential in order to make metamaterials' performance efficient.
However, resonance quality factors (that is the resonant frequency
over width of the resonance) demonstrated by conventional
metamaterials are often limited to rather small values.
This comes from the fact that resonating structural elements of metamaterials are strongly coupled to free-space
and therefore suffer significant losses due to radiation.
Furthermore, conventional metamaterials are often composed of
sub-wavelength particles that are simply unable to provide
large-volume confinement of electromagnetic field necessary to
support high-Q resonances. As recent theoretical analysis showed,
high-Q resonances involving dark (or closed) modes are nevertheless
possible in metamaterials if certain small asymmetries are
introduced in the shape of their structural elements
\cite{prosvirnin-2003-rcm}.

In this Letter we report the observation of exceptionally narrow
resonant responses in transmission and reflection of planar
metamaterial achieved through introducing asymmetry into its
structural elements. The appearance of narrow resonances is
attributed to the excitation of otherwise forbidden anti-symmetric
modes, that are weakly coupled to free-space ("dark modes"). 

Metamaterials that were used in our experiments consisted of
identical sub-wavelength metallic "inclusions" structured in the
form of asymmetrically split rings (ASR), which were arranged in a
periodic array and placed on a thin dielectric substrate (see
Fig.~1). ASR-patterns were etched from $35$~$\mu m$ copper
cladding covering IS620 PCB substrate of $1.5$~$mm$ thickness.
Each copper split ring had the radius of 6~$mm$ and width of
0.8~$mm$ and occupied a square translation cell of $15 \times
15$~$mm$ (see Fig. 1). Such periodic structure does not diffract
normal incident electromagnetic radiation for frequencies lower
than 20~$GHz$. The overall size of the samples used were
approximately $220 \times 220$~$mm$. Transmission and reflection
of a single sheet of this meta-material were measured in an
anechoic chamber under normal incidence conditions using broadband
horn antennas.

\begin{figure}[h]
\includegraphics[width=80mm]{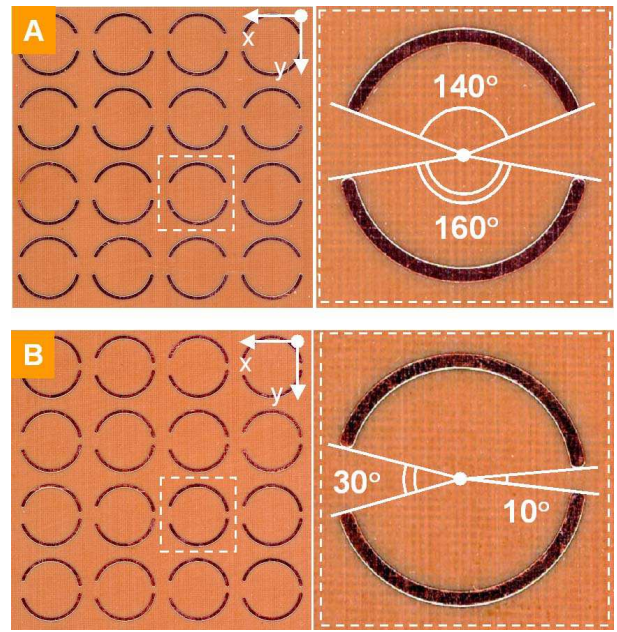}
\caption{(Color online) Fragments of planar metamaterials with
asymmetrically split copper rings.  The dashed boxes indicate
elementary translation cells of the structures.}
\end{figure}

We studied structures with two different types of asymmetry
designated as type A and B in Fig.~1. The rings of type A had two
equal splits dividing them into pairs of arcs of different length
corresponding to 140 and 160~$deg$ (see Fig.~1A). The rings of
type B were split along their diameter into two equal parts but
had splits of different length corresponding to 10 and 30~$deg$
(see Fig.~1B).

Transmission and reflection properties of structures of both types
depended strongly on the polarization state of incident
electromagnetic waves. The most dramatic spectral selectivity was
observed for electrical field being perpendicular to the mirror
line of the asymmetrically split rings, which corresponded to
$x$-polarization in the case of structure A and $y$-polarization
for structure B (as defined in Fig.~1). For the  orthogonal
polarizations the ASR-structures did not show any spectral
features originating from asymmetrical structuring.

\begin{figure}[h]
\includegraphics[width=80mm]{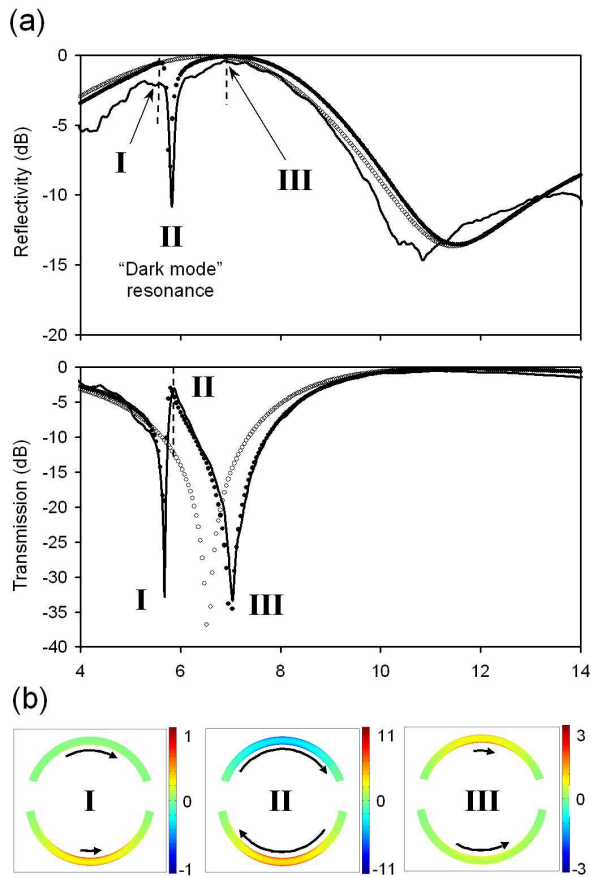}
\caption{(Color online) (a) Normal incidence reflection and
transmission spectra of A-type metamaterial (presented in Fig.~1A)
for $x$-polarization: solid line - experiment, filled circles -
theory (method of moments), empty circles - theory for reference
structure with symmetrically split rings. (b) $x$-Component of the
instantaneous current distribution in the asymmetrically split
rings corresponding to resonant features I, II and III as marked
in section (a). 
Arrows indicate instantaneous directions of the
current flow, while their length corresponds to the current
strength.}
\end{figure}

The results of reflection and transmission measurements of
metamaterial A obtained for $x$-polarization are presented in
Fig.~2a. The reflection spectrum reveals an ultra-sharp resonance
near 6~$GHz$ (marked as II), where reflectivity losses exceed
-10~$dB$. It is accompanied by two much weaker resonances (marked
as I and III) corresponding to reflection peaks at about 5.5 and
7.0~$GHz$ respectively. The sharp spectral response in reflection
is matched by a very narrow transmission peak reaching -3~$dB$ and
having the width of only 0.27~$GHz$ as measured at 3~$dB$ below
the maximum. The quality factor $Q$ of such response is 20, which
is larger than that of the most metamaterials based on lossy PCB substrates by at least one
order of magnitude. On both sides of the peak the transmission
decreases resonantly to about -35~$dB$ at frequencies
corresponding to reflection maxima.

Fig.~3a presents transmission and reflection spectra of B-type
metamaterial measured for $y$-polarization. A very narrow resonant
transmission dip can be seen near 5.5~$GHz$, where transmission
drops to about -5~$dB$. The corresponding reflection spectrum
shows an usually sharp roll-off (I-II) between -4 and -14~$dB$
spanning only 0.13~$GHz$ at around the same frequency. At the
frequency of about 11.5~$GHz$ the ASR-structure exhibits its
fundamental reflection resonance (marked as III) where the
wavelength of excitation becomes equal to the length of the arcs.

\begin{figure}[h]
\includegraphics[width=80mm]{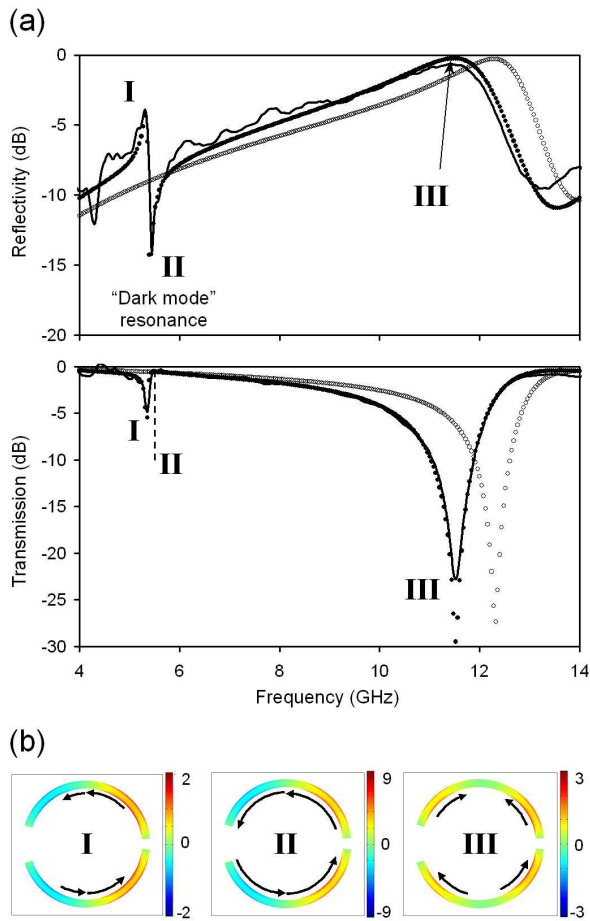}
\caption{(Color online) (a) Normal incidence reflection and
transmission spectra of B-type metamaterial (presented in Fig.~1B)
for $y$-polarization: solid line - experiment, filled circles -
theory (method of moments), empty circles - theory for reference
structure with symmetrically split rings. (b) $y$-Component of the
instantaneous current distribution in the asymmetrically split
rings corresponding to resonant features I, II and III as marked
in section (a). 
Arrows indicate instantaneous directions of the
current flow, while their length corresponds to the current
strength.}
\end{figure}

To understand the resonst nature of the response, the
ASR-structures were modelled using the method of moments. It is a
well established numerical method, which involves solving the integral equation for the surface current induced in the metal pattern by the field of the incident wave. This is followed by calculations of scattered fields as a superposition of partial spatial waves. The metal pattern is treated as a perfect conductor, while the substrate is assumed to be a lossy dielectric. For both
transmission and reflection the theoretical calculations show a
very good agreement with the experimental results assuming
$\epsilon = 4.07 + i\cdot 0.05$ (see Fig.~2 and 3, filled
circles). For comparison we also modelled metamaterial composed of
split rings with no structural asymmetry, i.e. equally split along
their diameter. Our calculations indicate that for both
polarizations the response of such structure is free form sharp
high-Q resonant feature (see Fig.~2 and 3, open circles).

The origin of the unusually strong and narrow spectral responses
of the ASR-structures can be traced to so-called "dark modes",
i.e. electromagnetic modes that are weakly coupled to free-space.
It is this property of the dark modes that allows in principal to
achieve high quality resonances in very thin structures
\cite{prosvirnin-2003-rcm}. These modes are usually forbidden but
can be excited in a planar metamaterial if, for example, its
particles have certain structural asymmetry.

Our calculations showed that in the case of structure A an
anti-symmetric current mode can dominate the usual symmetric one:
at the high-Q transmission resonance, as shown in Fig.~2b~(II),
two parts of the ring are excited in anti-phase while currents
have almost the same amplitude. The scattered electromagnetic fields produced by such current configuration are very weak, which dramatically reduces coupling to free-space and therefore radiation losses. As a consequence, the strength of the induced currents can reach very high values and therefore ensures high quality factor of the response. At the reflection resonances, in contrast to the
"dark mode" regime, currents in both sections of the
asymmetrically  split ring oscillate in phase but excitation of
one of the sections dominates the other (see Fig.~2b~(I and III)).
Importantly, the amplitudes of the currents in this case are
significantly smaller than in "dark mode" resonance, which yields
lower Q-factors for this type of the response. If the structural
asymmetry is removed the anti-symmetric current mode becomes
forbidden while two reflection resonances degenerate to a single
low-Q resonance state where both parts of the ring are excited
equally. Thus introduction of asymmetry in the split-ring
structure effectively allows to create a very narrow pass-band
inside its transmission stop-band. This effect is somewhat
analogous to appearance of an allowed state in the bandgap of
photonic crystals due to structural defects. 

Interpretation of the results obtained for structure B appears to be slightly more elaborate. From the symmetry of the split rings it follows that for $y$-polarized excitation at any frequency current distribution in the opposite sections of the ring should have equal $y$-components oscillating in phase and equal $x$-components oscillating in anti-phase. The net current in the ring has therefore always zero $x$-component, while its $y$-component can not be fully compensated due to the structural asymmetry. At low frequencies the net $y$-component is small but it increases significantly as the frequency of excitation approaches $~5.5$~GHz. At this frequency the wavelength becomes equal to circumference of the split ring and, as shown in Fig.~3b~(I), the right side of the ring dominates its left side oscillating in anti-phase. The later results in a resonant increase of the metamaterial reflection (see Fig.~3a). Immediately above this resonance contributions of both sides of the ASR-particle are still in anti-phase but become nearly identical (see Fig.~3b~(II)) making the $y$-component of the induced net current almost zero and therefore dramatically reducing radiation losses (reflection). Further increase of the excitation frequency leads to rise of the reflection until the fundamental resonance of the ASR-structure is reached where the corresponding wavelength is equal to the length of the arcs. In this case both sides of the ASR-particle oscillate in phase and equally contribute to electromagnetic field scattering (see Fig.~3b~(III)).

The quality factor of the dark-mode resonances will increase on reducing the degree of asymmetry of metamaterial particles and in case of low dissipative losses can be made exceptionally high. In the microwave region metals are almost perfect conductors and the main source of dissipative losses is the substrate material (dielectrics). Therefore significantly higher resonance quality factors can be achieved for a free-standing thin metal film, which is patterned complimentary to ASR-structure, i.e. periodically perforated with ASR-openings. In the visible and IR spectral ranges, however, losses in metals dominate and therefore nano-scaled versions of the original metal-dielectric ASR-structures would perform better. According to our estimates Q-factor of such ASR-nanostructures in the near-IR can be as high as 6. 

In summary, we experimentally and theoretically showed that a new
type of planar metamaterials composed of asymmetrically split
rings exhibit unusually strong high-Q resonances and provide for
extremely narrow transmission and reflection pass- and stop-bands.
The metamaterials' response has a quality factor of about 20,
which is one order of magnitude larger than the typical value for
many conventional metamaterials. This is achieved via weak
coupling between "dark modes" in the resonant inclusions of the
ASR-metamaterial and free-space, while weak symmetry breaking
enables excitation of so-called "dark modes". Achieving the "dark
mode" resonances will be especially important for metamaterials in
the optical part of the spectrum, where losses are significant and
unavoidable. In a certain way such symmetry-breaking resonances in
meta-materials resembles the recently identified spectral lines of
plasmon absorbtion of shell nanoparticle appearing due to
asymmetry \cite{Nanoegg}.

\begin{acknowledgments}
The authors would like to acknowledge the financial support of the
EPSRC (UK) and Metamorphose NoE.

\end{acknowledgments}


\begin{thebibliography}{8}
\expandafter\ifx\csname natexlab\endcsname\relax\def\natexlab#1{#1}\fi
\expandafter\ifx\csname bibnamefont\endcsname\relax
  \def\bibnamefont#1{#1}\fi
\expandafter\ifx\csname bibfnamefont\endcsname\relax
  \def\bibfnamefont#1{#1}\fi
\expandafter\ifx\csname citenamefont\endcsname\relax
  \def\citenamefont#1{#1}\fi
\expandafter\ifx\csname url\endcsname\relax
  \def\url#1{\texttt{#1}}\fi
\expandafter\ifx\csname urlprefix\endcsname\relax\def\urlprefix{URL }\fi
\providecommand{\bibinfo}[2]{#2}
\providecommand{\eprint}[2][]{\url{#2}}

\bibitem[{\citenamefont{Barnes et~al.}(2003)\citenamefont{Barnes, Dereux, and
  Ebbesen}}]{Ebessen}
\bibinfo{author}{\bibfnamefont{W.~L.} \bibnamefont{Barnes}},
  \bibinfo{author}{\bibfnamefont{A.}~\bibnamefont{Dereux}}, \bibnamefont{and}
  \bibinfo{author}{\bibfnamefont{T.~W.} \bibnamefont{Ebbesen}},
  \bibinfo{journal}{Nature} \textbf{\bibinfo{volume}{424}},
  \bibinfo{pages}{824} (\bibinfo{year}{2003}).

\bibitem[{\citenamefont{Smith et~al.}(2004)\citenamefont{Smith, Pendry, and
  Wiltshire}}]{NR}
\bibinfo{author}{\bibfnamefont{D.~R.} \bibnamefont{Smith}},
  \bibinfo{author}{\bibfnamefont{J.~B.} \bibnamefont{Pendry}},
  \bibnamefont{and} \bibinfo{author}{\bibfnamefont{M.~C.~K.}
  \bibnamefont{Wiltshire}}, \bibinfo{journal}{Science}
  \textbf{\bibinfo{volume}{305}}, \bibinfo{pages}{788} (\bibinfo{year}{2004}).

\bibitem[{\citenamefont{Fedotov et~al.}(2005)\citenamefont{Fedotov, Mladyonov,
  Prosvirnin, and Zheludev}}]{FS}
\bibinfo{author}{\bibfnamefont{V.~A.} \bibnamefont{Fedotov}},
  \bibinfo{author}{\bibfnamefont{P.~L.} \bibnamefont{Mladyonov}},
  \bibinfo{author}{\bibfnamefont{S.~L.} \bibnamefont{Prosvirnin}},
  \bibnamefont{and} \bibinfo{author}{\bibfnamefont{N.~I.}
  \bibnamefont{Zheludev}}, \bibinfo{journal}{Phys. Rev. E}
  \textbf{\bibinfo{volume}{72}}, \bibinfo{pages}{056613}
  (\bibinfo{year}{2005}).

\bibitem[{\citenamefont{Schwanecke et~al.}(2007)\citenamefont{Schwanecke,
  Fedotov, Khardikov, Prosvirnin, Chen, and Zheludev}}]{Mirror2}
\bibinfo{author}{\bibfnamefont{A.~S.} \bibnamefont{Schwanecke}},
  \bibinfo{author}{\bibfnamefont{V.~A.} \bibnamefont{Fedotov}},
  \bibinfo{author}{\bibfnamefont{V.}~\bibnamefont{Khardikov}},
  \bibinfo{author}{\bibfnamefont{S.~L.} \bibnamefont{Prosvirnin}},
  \bibinfo{author}{\bibfnamefont{Y.}~\bibnamefont{Chen}}, \bibnamefont{and}
  \bibinfo{author}{\bibfnamefont{N.~I.} \bibnamefont{Zheludev}},
  \bibinfo{journal}{J. Opt. A} \textbf{\bibinfo{volume}{9}},
  \bibinfo{pages}{L01} (\bibinfo{year}{2007}).

\bibitem[{\citenamefont{Fedotov et~al.}(2006)\citenamefont{Fedotov, Mladyonov,
  Prosvirnin, Rogacheva, Chen, and Zheludev}}]{Asymmetry}
\bibinfo{author}{\bibfnamefont{V.~A.} \bibnamefont{Fedotov}},
  \bibinfo{author}{\bibfnamefont{P.~L.} \bibnamefont{Mladyonov}},
  \bibinfo{author}{\bibfnamefont{S.~L.} \bibnamefont{Prosvirnin}},
  \bibinfo{author}{\bibfnamefont{A.~V.} \bibnamefont{Rogacheva}},
  \bibinfo{author}{\bibfnamefont{Y.}~\bibnamefont{Chen}}, \bibnamefont{and}
  \bibinfo{author}{\bibfnamefont{N.~I.} \bibnamefont{Zheludev}},
  \bibinfo{journal}{Phys. Rev. Lett.} \textbf{\bibinfo{volume}{97}},
  \bibinfo{pages}{167401} (\bibinfo{year}{2006}).

\bibitem[{\citenamefont{Schurig et~al.}(2006)\citenamefont{Schurig, Mock,
  Justice, Cummer, Pendry, Starr, and Smith}}]{Cloak}
\bibinfo{author}{\bibfnamefont{D.}~\bibnamefont{Schurig}},
  \bibinfo{author}{\bibfnamefont{J.~J.} \bibnamefont{Mock}},
  \bibinfo{author}{\bibfnamefont{B.~J.} \bibnamefont{Justice}},
  \bibinfo{author}{\bibfnamefont{S.~A.} \bibnamefont{Cummer}},
  \bibinfo{author}{\bibfnamefont{J.~B.} \bibnamefont{Pendry}},
  \bibinfo{author}{\bibfnamefont{A.~F.} \bibnamefont{Starr}}, \bibnamefont{and}
  \bibinfo{author}{\bibfnamefont{D.~R.} \bibnamefont{Smith}},
  \bibinfo{journal}{Science} \textbf{\bibinfo{volume}{314}},
  \bibinfo{pages}{977} (\bibinfo{year}{2006}).

\bibitem[{\citenamefont{Prosvirnin and Zouhdi}(2003)}]{prosvirnin-2003-rcm}
\bibinfo{author}{\bibfnamefont{S.}~\bibnamefont{Prosvirnin}} \bibnamefont{and}
  \bibinfo{author}{\bibfnamefont{S.}~\bibnamefont{Zouhdi}}, in
  \emph{\bibinfo{booktitle}{Advances in Electromagnetics of Complex Media and
  Metamaterials}}, edited by
  \bibinfo{editor}{\bibfnamefont{S.}~\bibnamefont{Zouhdi}} \bibnamefont{and}
  \bibinfo{editor}{\bibnamefont{et~al.}} (\bibinfo{publisher}{Kluwer Academic
  Publishers}, \bibinfo{address}{Printed in the Netherlands},
  \bibinfo{year}{2003}), pp. \bibinfo{pages}{281--290}.

\bibitem[{\citenamefont{Wang et~al.}(2006)\citenamefont{Wang, Wu, Lassiter,
  Nehl, Hafner, Nordlander, and Halas}}]{Nanoegg}
\bibinfo{author}{\bibfnamefont{H.}~\bibnamefont{Wang}},
  \bibinfo{author}{\bibfnamefont{Y.}~\bibnamefont{Wu}},
  \bibinfo{author}{\bibfnamefont{B.}~\bibnamefont{Lassiter}},
  \bibinfo{author}{\bibfnamefont{C.~L.} \bibnamefont{Nehl}},
  \bibinfo{author}{\bibfnamefont{J.~H.} \bibnamefont{Hafner}},
  \bibinfo{author}{\bibfnamefont{P.}~\bibnamefont{Nordlander}},
  \bibnamefont{and} \bibinfo{author}{\bibfnamefont{N.~J.} \bibnamefont{Halas}},
  in \emph{\bibinfo{booktitle}{Proceedings of the National Academy of Science
  of the United States of America}} (\bibinfo{year}{2006}), vol.
  \bibinfo{volume}{103}, p. \bibinfo{pages}{10856}.

\end{thebibliography}
\end{document}